\documentclass[preprint,notoc]{JHEP3}
\usepackage{epsfig}
\def\fa{f_a}

\def\to{\rightarrow}

\def\bi{\begin{itemize}}
\def\ei{\end{itemize}}

\def\c1p{C1^\prime}
\def\ta{\tilde a}
\def\tG{\widetilde G}

\def\ta{\tilde a}

\def\tg{\tilde g}

\def\tz{\widetilde Z}

\def\sigv{\langle \sigma v \rangle}
\def\To{\Rightarrow}
\def\alt{\lesssim}
\def\agt{\gtrsim}
\def\be{\begin{equation}}  
\def\ee{\end{equation}}  
\def\bea{\begin{eqnarray}}  
\def\eea{\end{eqnarray}}

\newcommand\njp[3]{{\it New\ J.\ Phys.\ }{\bf #1} (#2) #3}

\newcommand\sjp[3]{{\it Sov.\ J.\ Nucl.\ }{\bf #1} (#2) #3}

\def\Isajet{{\sc Isajet}}

\preprint{\vbox{OU-HEP-111006}}

\title{Coupled Boltzmann calculation of\\
mixed axion/neutralino cold dark matter\\
production in the early universe
}
\author{Howard Baer$^{a}$, Andre Lessa$^{b}$ and Warintorn Sreethawong$^a$\\
$^a$Dept.\ of Physics and Astronomy, University of Oklahoma, Norman, OK 73019, USA\\
$^b$ Instituto de F\'isica, Universidade de S\~ao Paulo, S\~ao Paulo - SP, Brazil\\
E-mail: \email{baer@nhn.ou.edu}, \email{lessa@fma.if.usp.br}, 
\email{wstan@nhn.ou.edu}}

\abstract{
We calculate the relic abundance of mixed axion/neutralino cold dark matter 
which arises in $R$-parity conserving supersymmetric (SUSY) models 
wherein the strong CP problem is solved by 
the Peccei-Quinn (PQ) mechanism with a concommitant axion/saxion/axino 
supermultiplet.
By numerically solving the coupled Boltzmann equations, we include the combined effects of 
1. thermal axino production with cascade decays to a neutralino LSP, 
2. thermal saxion production and production via coherent 
oscillations along with cascade decays and entropy injection, 
3. thermal neutralino production and re-annihilation after both axino and saxion decays, 
4. gravitino production and decay and 
5. axion production both thermally and via oscillations. 
For SUSY models with too high a standard neutralino thermal abundance, 
we find the combined effect of SUSY PQ particles is not enough to lower the
neutralino abundance down to its measured value, while at the same time respecting
bounds on late-decaying neutral particles from BBN.
However, models with a standard neutralino underabundance can now be 
allowed with either neutralino or axion domination of dark matter, and furthermore,
these models can allow the PQ breaking scale $\fa$
to be pushed up into the $10^{14}-10^{15}$ GeV range, which is 
where it is typically expected to be in string theory models.
}

\begin{document}

\section{Introduction}
\label{sec:intro}

The Standard Model (SM) of particle physics is beset by two afflictions: 
1. in the scalar (Higgs) sector of the theory, quadratic divergences require large fine-tunings of 
electroweak parameters which depend on the scale $\Lambda$ below which the SM is regarded
as the correct effective field theory of nature and 2. in the QCD sector of the theory,
the Lagrangian term
\be
{\cal L}\ni \frac{\bar{\theta}}{32\pi^2}F_{A\mu\nu}\tilde{F}_A^{\mu\nu}
\label{eq:L_th}
\ee
required by 'tHooft's solution to the $U(1)_A$ problem is constrained to a value $\bar{\theta}\alt 10^{-10}$
to gain accord with measurements of the neutron EDM\cite{peccei}. 
The first of these is solved by 
the introduction of softly broken weak scale supersymmetry (SUSY) into the theory\cite{wss} 
(which receives some indirect support from the measured values of gauge couplings at LEP\cite{gcu}
and from global fits to precision electroweak data\cite{ew}), 
while the second problem is solved by the introduction of a global $U(1)_{PQ}$ Peccei-Quinn (PQ)
symmetry broken by QCD anomalies\cite{pqww}, 
which requires the existence of an (``invisible'') axion\cite{ksvz,dfsz}, 
with mass expected in the micro-eV or below range\cite{axreview}. 
Solving both problems simultaneously requires supersymmetrization of the 
SM (usually via the Minimal Supersymmetric Standard Model, or MSSM) 
along with the introduction of an axion supermultiplet $\hat{a}$ into the theory. 
The $\hat{a}$ supermultiplet contains an $R$-parity-even spin-0 saxion field $s(x)$
along with an $R$-parity-odd spin-${1\over 2}$ axino $\ta (x)$, in addition to the
usual pseudoscalar axion field $a(x)$:
\be
\hat{a}=\frac{s+ia}{\sqrt{2}}+ i\sqrt{2}\bar{\theta}\ta_L +i\bar{\theta}\theta_L{\cal F}_a ,
\ee
in 4-component spinor notation\cite{wss}.
 
In such a theory, it is expected that SM superpartner particles with 
weak scale masses should emerge, 
along with a weak scale saxion, whilst the axino mass is more model dependent, with 
$m_{\ta}\sim $ keV-TeV being expected\cite{axinorev}. 
The axion, saxion and axino couplings to matter
depend on the PQ breaking scale $\fa$\footnote{Throughout this work
we omit the number of generations factor $N$, which appear along with
the PQ scale, $f_a/N$, in the DSFZ model and in the KSVZ model with more than
one heavy quark generation. All our results can then be trivially generalized replacing
$f_a$ by $f_a/N$.}, which is required $\fa \agt 10^9$ GeV by
stellar cooling calculations\cite{astro}.
The axion is often considered as a very appealing dark matter (DM) candidate\cite{axiondm,sik}.
\footnote{For a somewhat different axion/axino scenario, see Ref. \cite{lyth}.}

In the MSSM, DM candidates include the lightest neutralino $\tz_1$ (a WIMP), 
the spin-$3\over 2$ gravitino $\tG$ or 
possibly the superpartner of a right-handed neutrino\cite{snu_R}. 
Gravitino dark matter is tightly constrained and disfavored by the standard picture of 
Big Bang nucleosynthesis (BBN)\cite{moroi}, whilst right-hand neutrino states are expected
to exist near the GUT scale according to the elegant see-saw mechanism for 
neutrino mass\cite{seesaw}.
Many authors thus expect dark matter to be comprised of the SUSY neutralinos, a natural WIMP
candidate which is motivated by the so-called ``WIMP miracle''. 
However, detailed analyses show that neutralino dark matter requires a rather 
high degree of fine-tuning\cite{ft} to match the WMAP-measured
cold DM abundance\cite{wmap7}:
\be
\Omega_{\rm DM}h^2=0.1123\pm 0.0035\ \ \ {\rm at\ 68\%\ CL} .
\ee
In fact, the measured abundance lies in the most improbable locus of values
of neutralino relic density as predicted by general scans over SUSY model parameter space\cite{bbs2}.

The PQ-extended Minimal Supersymmetric Standard Model (PQMSSM) offers additional
possibilities to describe the dark matter content of the universe. 
In the PQMSSM, the axino may play the role of stable lightest SUSY partner (LSP)\cite{rtw,ckkr}, 
while the quasi-stable axion may also constitute a component of DM\cite{bbs1}, giving rise 
to mixed axion/axino ($a\ta$) CDM.
In supergravity theories however, the axino mass is expected to lie at the weak scale\cite{cl}, 
so that the neutralino remains as LSP, and the possibility occurs for
mixed axion/neutralino ($a\tz_1$) CDM.

In a recent paper, Choi {\it et al.}\cite{ckls} presented a semi-analytic approach for
estimating the relic abundance of neutralinos in the mixed $a\tz_1$ CDM scenario.
This approach applies to cases where the thermally averaged neutralino annihilation cross
section times relative velocity $\langle\sigma v\rangle$ is approximately constant
with temperature, as occurs for a wino-like or higgsino-like neutralino\cite{z1dmreview}. 
Detailed calculations of the relic abundance of mixed $a\tz_1$ CDM were performed in
Ref. \cite{blrs}, where formulae for the neutralino and axion abundances were
presented. 

The standard calculation of the neutralino Yield 
$Y_{\tz_1}^{std}\equiv \frac{n_{\tz_1}}{s}$ (where $n_{\tz_1}$ is the neutralino number
density and $s$ is the entropy density) gives
\be
Y_{\tz_1}^{std}=\frac{\left(90/\pi^2g_*(T_{fr})\right)^{1/2}}{4\langle\sigma v\rangle M_PT_{fr}} ,
\label{eq:Ystd}
\ee
where $g_*(T_{fr})$ is the number of active degrees of freedom at temperature
$T=T_{fr}$, where
\be
T_{fr}^{std} = m_{\tz_1}/\ln[\frac{3\sqrt{5}\langle \sigma v\rangle M_P m_{\tz_1}^{3/2}}
{\pi^{5/2}T_{fr}^{1/2}g_*^{1/2}(T_{fr})}] .
\label{eq:tfrz1}
\ee
is the freeze-out temperature and $M_P$ is the reduced Planck mass. 

If instead axinos are thermally produced (TP) at a large rate at re-heat temperature $T_R$
after inflation, then they cascade decay to (stable) neutralinos at decay 
temperature 
\be
T_D^{\ta}=\sqrt{\Gamma_{\ta}M_P}/\left(\pi^2g_*(T_D^{\ta})/90\right)^{1/4} ,
\label{eq:TD}
\ee
and can boost the neutralino abundance. 
The late-time injection of neutralinos into the cosmic soup at temperatures $T_D^{\ta}<T_{fr}$ 
may cause a {\it neutralino re-annihilation effect}
such that the neutralino Yield is instead given by\cite{ckls,blrs}
\be
Y_{\tz_1}^{re-ann}|_{T=T_D^{\ta}}\simeq
\frac{\left(90/\pi^2g_*(T_D^{\ta})\right)^{1/2}}{4\langle\sigma v\rangle M_PT_D^{\ta}} .
\label{eq:Yreann}
\ee
Since $T_D^{\ta}$ is typically in the MeV-GeV range, {\it i.e.} well below
$T_{fr}\sim m_{\tz_1}/20$, the neutralino abundance after re-annihilation
can be highly enhanced relative to the standard cosmological picture.
In addition, one must fold into the relic abundance the axion contribution arising
from coherent axion field oscillations beginning at axion oscillation 
temperature $T_a\sim 1$ GeV.

An additional complication comes from entropy production from
axino decay after $T_{fr}$ (which may dilute the neutralino abundance) or
after $T_a$ (which may dilute the axion abundance). This may occur
in the case where axinos temporarily dominate the energy density of the universe.
Depending on the PQ parameters of the PQMSSM model 
($\fa,\ m_{\ta}$, initial axion misalignment angle $\theta_i$ and $T_R$), 
the dark matter abundance may be either neutralino- or axion-dominated. In fact,
cases may occur where the DM relic abundance is shared comparably between the two.
In the latter case, it might be possible to directly detect relic neutralino WIMP
particles as well as relic axions!

While the semi-analytic treatment of Ref's \cite{ckls} and \cite{blrs} provides a broad
portrait of the mixed $a\tz_1$ CDM picture, a number of important features have
been neglected. These include the following.
\bi
\item For bino-like neutralinos, $\langle\sigma v\rangle\sim a+bT^2$ where $a\sim 0$ 
since we mainly have $p$-wave annihilation cross sections. In this case,
$\langle\sigma v\rangle$ is no longer independent of temperature, and 
the simple formulae \ref{eq:Ystd} and \ref{eq:Yreann} are no longer valid.
\item In Ref's \cite{ckls} and \cite{blrs}, the effects of saxion production and
decay in the early universe are neglected. In fact, saxion thermal production or
production via coherent oscillations (CO)\cite{turnerosc}, followed by late time saxion decay, may
inject considerable entropy into the early universe, thus diluting all relics
present at the saxion decay temperature $T_D^{s}$. Saxions may also add to the neutralino
abundance via decays such as $s\to\tg\tg$, followed by gluino cascade decays.
There exists the possibility of saxion and axino {\it co-domination} of the
universe.
In this case, there might be a second neutralino re-annihilation taking place at $T_D^{s}$.
\item The treatments of \cite{ckls} and \cite{blrs} invoke the ``sudden decay''
approximation for late-decaying axinos, whereas in fact the decay process is a continuous
one proceeding in time until the decaying species is highly depleted (all have decayed).
\item The treatments of \cite{ckls} and \cite{blrs} largely ignore the effect of gravitino
production and decay in the early universe.
\ei

To include the above effects into a calculation of the mixed $a\tz_1$ relic abundance, 
one must go beyond the semi-analytic treatment presented in Ref's \cite{ckls,blrs}, and
proceed with a full solution of the coupled Boltzmann equations which govern 
various abundances of neutralinos, axinos, axions, saxions, gravitinos and radiation.

Toward this end, in Sec. \ref{sec:calc} we present a simplified set of coupled Boltzmann equations,
which we use to calculate the relic abundance of mixed axion/neutralino dark matter. 
More details about the approximations made and each term present in our equations are
discussed in Appendix \ref{sec:boltzeqs}.

In Sec. \ref{sec:numerics}, we 
present various numerical results for the
mixed $a\tz_1$ CDM scenario using the full set of Boltzmann equations.
We find that, even after the inclusion of the saxion field, 
adjusting the parameters of the PQMSSM can only {\it increase} the neutralino abundance,
 and not decrease
it, while at the same time respecting bounds on late-decaying neutral particles from BBN.
This result is the same as found in Refs. \cite{ckls} and \cite{blrs}, but now
corresponds to a much stronger statement, since the saxion entropy injection
had been neglected in the previous works. Furthermore, our results also apply
to models with bino-like neutralinos, which could not be studied in the semi-analytical
framework used in Refs. \cite{ckls} and \cite{blrs}.

Since the neutralino abundance can be only enhanced in the PQMSSM, in models such as mSUGRA,
those points which are excluded by a {\it standard overabundance} of neutralinos are
still excluded in the PQMSSM!
This rather strong conclusion does depend on at least three assumptions: 1. that thermal axino 
production rates are not suppressed by low-lying $PQ$-charged matter multiplets\cite{bci}\footnote{
Here, we assume standard rates for thermal axino production as calculated in the 
literature\cite{ckkr,graf,strumia}.
In Ref. \cite{bci}, it has been shown that if $PQ$-charged matter multiplets $\hat{\Phi}$ exist well below the PQ
breaking scale $\fa$, then axino production is suppressed by factors of $m_{\Phi}/T_R$.
}, 2. that saxion
decay is dominated by gluon and gluino pairs and 3. that the assumed saxion field strength 
$s(x)\equiv \theta_s\fa$ is of order the PQ-breaking scale $\fa$, {\it i.e.} that $\theta_s\sim 1$.

We also examine several cases with a {\it standard underabundance} of neutralino dark matter.
In these cases, again the neutralino abundance is only increased (if BBN constraints
are respected). Thus, adjustment of PQMSSM parameters can bring models with an underabundance of neutralinos
into accord with the measured DM relic density. In these cases, the DM abundance 
tends to be neutralino-dominated. 
Also, in these cases, solutions exist where the PQ scale $\fa$ is either near its lower range, or 
where $\fa$ is much closer to $M_{GUT}$, with $\fa\sim 10^{14}$ GeV typically allowed.
This is much closer to the scale of $\fa$ which is thought to arise from 
string theory\cite{witten}.
In Sec. \ref{sec:conclude}, we present a summary and conclusions.

\section{Mixed axion/neutralino abundance from coupled Boltzmann equations}
\label{sec:calc}

Here, we present a brief description of our procedure to calculate the
relic abundance of mixed $a\tz_1$ CDM in the PQMSSM. A more detailed discussion
is left to Appendix \ref{sec:boltzeqs}.

\subsection{Boltzmann equations}

The general Boltzmann equation for the number density of a particle species can be generically
 written as\cite{kt}:
\be
\dot{n}_i + 3 H n_i = S_i - \frac{1}{\gamma_i} \Gamma_i n_i
\ee
where $S_i$ represents a source term, $\Gamma_i$ is the decay width and $\gamma_i$ is the relativistic dilation factor
to take into account the suppressed decays of relativistic particles.
To describe the thermal production of a particle species $i$ as well as its decoupling
from the radiation fluid and the non-thermal production coming from other particles decays,
we include in $S_i$ the following terms:
\be
S_i = -[n_{i}^2 - (n^{eq}_{i}(T))^2] \sigv_i(T) + \sum_{j} BR(j,i) \Gamma_j \frac{n_j}{\gamma_j}
\ee
where $\langle\sigma v\rangle$ is the  (temperature dependent) thermally averaged annihilation cross section times velocity for the particle species $i$, $n^{eq}_i$ is its equilibrium number density and
$BR(j,i)$ is the branching fraction for particle $j$ to decay to particle $i$.\footnote{In this paper, $i$ is summed over
1. neutralinos $\tz_1$, 2. TP axinos $\ta$, 3. and 4. CO- and TP-produced saxions $s(x)$, 5. and 6. CO- and TP-axions $a$, 
7. TP gravitinos $\tG$ and radiation. We allow for axino decay to $g\tg$, $\gamma\tz_i$ and $Z\tz_i$ states ($i=1-4$), and
saxion decay to $gg$ and $\tg\tg$. Additional model-dependent saxion decays {\it e.g.} to $aa$ and/or $hh$ are
possible and would modify our results. We assume $\tG$ decay to all particle-sparticle pairs, and include 3-body gravitino 
modes as well\cite{moroi_grav}.}

The Boltzmann equation then becomes:
\be
\dot{n}_i + 3 H n_i = - \Gamma_i m_i \frac{n_i^2}{\rho_i} + [(n^{eq}_{i}(T))^2 - n_{i}^2] \sigv_i + \sum_{j} BR(j,i) \Gamma_j m_j \frac{n_j^2}{\rho_j} \; ,
 \label{nieq}
\ee
where we have used $\gamma_i = \rho_i/m_i n_i$. As discussed in Appendix \ref{sec:boltzeqs},
the above equation is also valid for coherent oscillating fields once we take $BR(j,i)=0$ and
$\sigv_i = 0$.

It is also convenient to write an equation for the evolution of entropy:
\bea
\dot{S} & = &\left(\frac{2 \pi^2}{45} g_*(T) \frac{1}{S}\right)^{1/3} R^4 \sum_{i} BR(i,X) \frac{1}{\gamma_i} \Gamma_i \rho_{i} \nonumber \\
{\rm or}\ \ \  \dot{S} & = &\frac{R^3}{T} \sum_{i} BR(i,X) \Gamma_i m_i n_{i} 
\label{Seq}
\eea
where $BR(i,X)$ is the fraction of energy injected in the thermal bath from $i$ decays.
 
Along with Friedmann's equation,
\be
H = \frac{1}{R} \frac{dR}{dt} = \sqrt{\frac{\rho_T}{3 M_P^2}} \; , \mbox{ with } \rho_T \equiv \sum_{i} \rho_i + \frac{\pi^2}{30} g_*(T) T^4 \; ,
\label{H}
\ee
the set of coupled differential equations, Eq's.~\ref{nieq}, \ref{Seq} and \ref{H},
can be solved as a function of time. More details on the solution of the above equations
and the expressions used for $\sigv_i$, $BR(i,j)$ and $BR(i,X)$ are presented in
Appendix \ref{sec:boltzeqs}.

\subsection{Present day abundances and constraints from BBN} 
\label{ssec:bbn}

To compute the relic density of neutralinos and axions we evolve the various particle and sparticle abundances from $T=T_R$ until the final temperature
$T_F$ is reached at which all unstable particles (save the axion itself) have decayed.
The relic densities of the various dark matter species labeled by $i$ are then given by:
\be
\Omega_i h^2 = \frac{\rho_i(T_F)}{s(T_F)} \times \frac{s(T_{CMB})}{\rho_c/h^2}.
\ee

In our calculations, a critical constraint comes from maintaining the
success of the standard picture of Big Bang nucleosynthesis. Constraints from BBN 
on late decaying neutral particles (labeled $X$) have been
calculated recently by several groups\cite{ellis,kohri,jedamzik} (we explicitly use the 
results of Ref. \cite{jedamzik}) 
and are presented as functions of
1. the decaying neutral particle's hadronic branching fraction $B_h$, 2. the decaying particle's
lifetime $\tau_X$, and 3. the decaying particle's relic abundance $\Omega_Xh^2$ had it not decayed.
The constraints also depend on 4. the decaying particle's mass $m_X$. We have constructed
digitized fits to the constraints given in Ref. \cite{jedamzik}, and apply these to late decaying
gravitinos, axinos and saxions. Typically, unstable neutrals with decay temperature below
5~MeV (decaying during or after BBN) and/or large abundances will be more likely to destroy
the predicted light element abundances.

\subsection{Example: calculation from a generic mSUGRA point} 
\label{ssec:example}

As an example calculation, we adopt a benchmark point from the paradigm minimal
supergravity model (mSUGRA), with parameters $(m_0,\ m_{1/2},\ A_0,\ \tan\beta,\ sign(\mu ))=$
(400~GeV, 400~GeV, 0, 10, +). The sparticle mass spectrum is generated by Isasugra\cite{isasugra},
and has a bino-like neutralino with mass $m_{\tz_1}=162.9$ GeV and a standard relic abundance
from IsaReD\cite{isared} of $\Omega_{\tz_1}^{std}h^2=1.9$ 
(it would thus be excluded by WMAP7 measurements assuming the standard neutralino freeze-out calculation).
We assume a gravitino mass $m_{\tG}=1$ TeV.

\begin{figure}[t]
\begin{center}
\includegraphics[width=14cm]{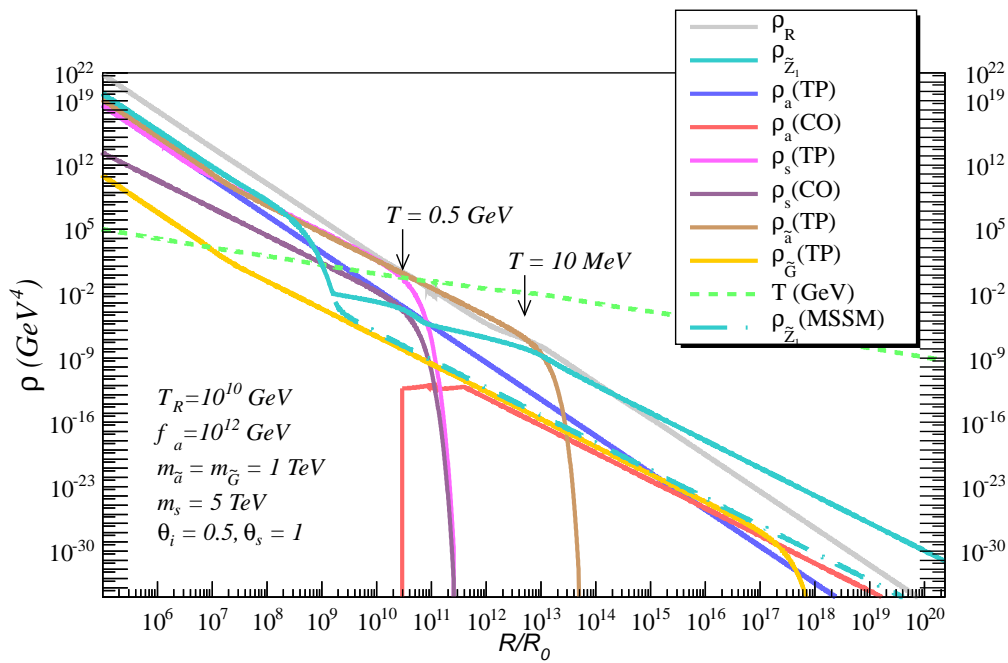}
\caption{Evolution of radiation, neutralino, axion, saxion, axino and gravitino
energy densities versus scale factor $R$.
We adopt an mSUGRA SUSY model with parameters
$(m_0,m_{1/2},A_0,\tan\beta,sign(\mu ))=(400\ {\rm GeV},400\ {\rm GeV}, 0,10,+)$.
We also take $m_{\tG}=1$ TeV and $T_R=10^{10}$ GeV and PQ parameters
$m_{\ta}=1$ TeV, $m_s=5$~TeV, $\theta_i=0.5$, $\theta_s=1$
with $\fa=10^{12}$ GeV.
}
\label{fig:bench}
\end{center}
\end{figure}

Here, we work in the PQMSSM framework, and take $T_R=10^{10}$ GeV with PQ parameters as
$m_{\ta}=1$ TeV, $m_s=5$ TeV, $\theta_i=0.5$ and $\fa =10^{12}$ GeV. 
We also take $\theta_s = 1$, where $\theta_s f_a$ is the initial field amplitude for
coherent oscillating saxions.
The various energy densities $\rho_i$ are shown in Fig.  \ref{fig:bench}
for $i=R$ (radiation), $\tz_1$ (neutralinos), $a^{TP}$ (thermally produced axions), $a^{CO}$ (coherent oscillating axions), $s^{TP}$ (thermally produced saxions), $s^{CO}$ (coherent oscillating
saxions), $\ta^{TP}$ (thermally produced axinos)
and $\tG^{TP}$ (thermally produced gravitinos). The energy densities are plotted against scale factor ratio $R/R_0$, 
where $R_0$ is the scale factor at $T=T_R$. We also plot the temperature $T$ of
radiation (green-dashed curve). 

We see that, at $R/R_0 < 10^{10}$, the universe is indeed radiation-dominated.
At $T \gg 1$ TeV, the TP axions, saxions and axinos all have similar abundances. At
these temperatures, the saxion coherent abundance as well as the gravitino
thermal abundance are far below the other components. As the universe expands and cools, most components
are relativistic, and decrease with the same slope as radiation: $\rho_i\sim T^{-4}$. The exception 
is the CO-produced saxions, which are non-relativistic, and fall-off as $\rho_{s}^{CO}\sim T^{-3}$.
 At $R/R_0\sim 10^7$, the
temperature $T\sim 1$ TeV, and the thermally-produced axinos, saxions and gravitinos become non-relativistic, 
so now $\rho_{\ta,s,\tG}^{TP}\sim T^{-3}$. For even lower temperatures with
$R/R_0\sim 10^9$, neutralinos begin to freeze-out, and their abundance falls steeply.
At $T\sim m_{\tz_1}/20$, they do freeze-out, and normally their density would fall as
$\rho_{\tz_1}\sim T^{-3}$, as indicated by the blue dot-dashed curve, which shows neutralino
abundance in the MSSM, without PQ-augmentation. 
In the PQMSSM however, saxions-- and later still axinos-- begin decaying in earnest,
and feed into the neutralino abundance, preventing its usual fall as $T^{-3}$.
At $T \sim 0.5$ GeV, the energy density of axinos surpass the radiation component and the
universe becomes axino-dominated until the axino decays at $T \sim 10$ MeV.
Also, around $R/R_0\sim 3\times 10^{10}$ with $T\sim 1$ GeV, CO production of axions begins, and by
$R/R_0\sim 4\times 10^{11}$, with $T\lesssim \Lambda_{QCD}$, its abundance begins to fall as $T^{-3}$. For even lower temperatures ($T < 10$ MeV), 
the axinos have essentially all decayed, feeding back into the neutralino abundance, and
also increasing the entropy per co-moving volume, which would otherwise be conserved.
At $R/R_0\sim 10^{14}$, the universe moves from radiation domination to matter (neutralino)
domination, while at even lower temperatures, the gravitinos decay away.
In this case, the final neutralino abundance is $\Omega_{\tz_1}h^2\sim 90017$-- far beyond its
standard value. This is mainly due to its abundance being augmented by thermal axino and
saxion production and 
cascade decay to neutralinos. In the standard
axion cosmology, the axion abundance would have been $\Omega_a^{std}h^2\sim 0.06$\cite{vg1}. 
In the case illustrated here,
entropy injection from saxion, axino and gravitino decays has diluted its abundance to just
$\Omega_ah^2\sim 0.004$.

%

\begin{figure}[t]
\begin{center}
\includegraphics[width=14cm]{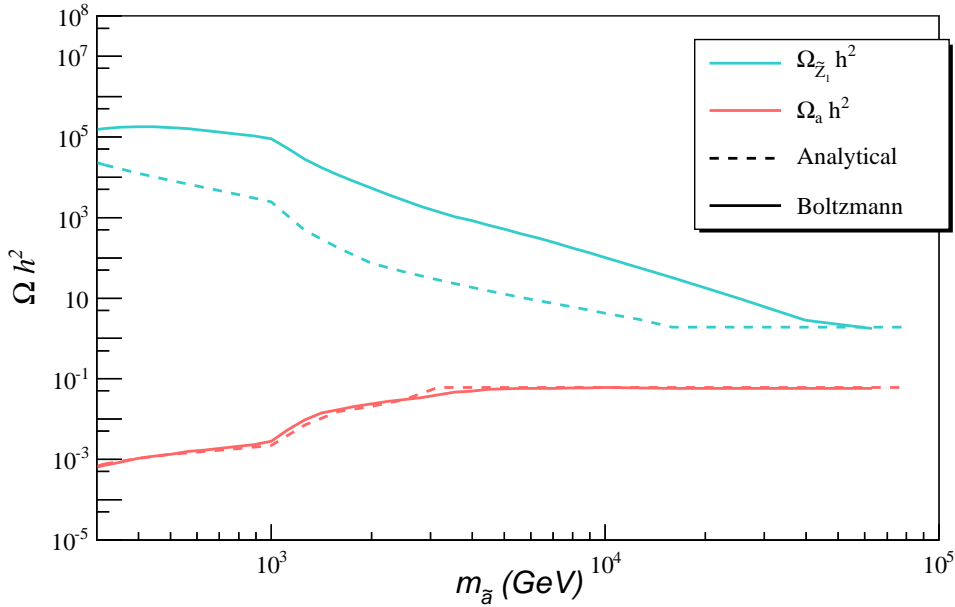}
\caption{Neutralino and axion relic densities as a function of the axino mass
for $\theta_i=0.5$, $T_R = 10^{10}$ GeV, $\fa=10^{12}$ GeV and the mSUGRA
point $(m_0,m_{1/2},A_0,\tan\beta,sign(\mu ))=(400\ {\rm GeV},400\ {\rm GeV}, 0,10,+)$.
The solid lines correspond to the solution of the Boltzmann equations while
the dashed lines correspond to the results obtained using the analytical expressions
derived in Ref. \cite{blrs}.}
\label{fig:omvsmax}
\end{center}
\end{figure}

As an example of the relevance of using the full set of Boltzmann equations
instead of the semi-analytical approach of Refs. \cite{ckls} and \cite{blrs},
we compare in Fig.  \ref{fig:omvsmax} the neutralino and axion relic densities
as a function of the axino mass using the Boltzmann equation formalism
and the semi-analytical approach. The other PQMSSM parameters are the same as used in Fig.  \ref{fig:bench},
but to compare with the semi-analytical results of Refs. \cite{ckls} and \cite{blrs} we
neglect the saxion component.
For these choices of PQ parameters and for $m_{\ta} \lesssim 50$~TeV, the axino decays after neutralino
freeze-out (as seen on Fig.  \ref{fig:bench}), significantly enhancing its final relic abundance.
Furthermore, the axino decay injects entropy, diluting the axion abundance.
As we can see from Fig.  \ref{fig:omvsmax}, the axion relic density obtained using
the analytical expressions derived in Ref. \cite{blrs} agree extremely well with
the solution of the Boltzmann equations. On the other hand, the analytic neutralino
abundance disagrees with the Boltzmann solution by almost an
order of magnitude for $m_{\ta} \lesssim 50$~TeV.
The primary reason for this is the fact that, for this mSUGRA point,
the neutralino is bino-like and $\sigv_{\tz_1}$ is no longer constant, but strongly
depends on the temperature. This dependence has not been included in the
semi-analytical approach. Also, the sharp transition seen in the semi-analytical result at 
$m_{\ta} \simeq 18$~TeV, where $T_D^{\ta}$ becomes bigger than $T_{fr}$, is artificially
introduced by the sudden decay approximation. As shown by the Boltzmann solution, the enhancement
of the neutralino relic abundance smoothly decreases, going up to $m_{\ta} \simeq 50$~TeV.

%

\section{Neutralino abundance in the PQMSSM}
\label{sec:numerics}
\subsection{Neutralino Abundance in several PQMSSM models}

In this section, we adopt four SUSY benchmark models listed in Table \ref{tab:bm}. 
The first two points, labeled BM1 and BM2, are generic mSUGRA points with
a bino-like $\tz_1$ which give rise as expected to an apparent excess of CDM.
For BM1, with $(m_0$, $m_{1/2}$, $A_0$, $\tan\beta$, $sign(\mu))=$ (400~GeV, 400~GeV, 0, 10, +)
we have $m_{\tz_1}=162.8$ GeV with a standard abundance $\Omega_{\tz_1}^{std}h^2=1.9$, while 
the second point (BM2) has 
$(m_0$, $m_{1/2}$, $A_0$, $\tan\beta$, $sign(\mu))=$ (3000~GeV, 1000~GeV, 0, 10, +)
with $m_{\tz_1}=436.3$ GeV and $\Omega_{\tz_1}^{std}h^2=49.6$. 
The next point, BM3, is a mSUGRA point with an apparent underabundance of neutralino
dark matter, with $m_{\tz_1}=163.8$ GeV, $m_A= 367.5$ GeV, lying in the $A$-funnel region\cite{Afunnel},
so $\Omega_{\tz_1}^{std}h^2=0.019$.
For all cases, we take $m_{\tG}=1$ TeV, 
but now will vary the PQ parameters and $T_R$, in order to see if 
the relic density of mixed $a\tz_1$ CDM can lie in the WMAP-allowed region.
The last point is taken from the gaugino AMSB model\cite{amsb,inoAMSB} and has a
wino-like neutralino with $\Omega_{\tz_1}^{std}h^2=0.0016$, but with
$m_{3/2}\equiv m_{\tG}=50$ TeV.
%
\begin{table}\centering
\begin{tabular}{lccccc}
\hline
 & BM1  & BM2 & BM3 & BM4 \\
\hline
$m_0$ & 400 & 3000 & 400 & 0 \\
$m_{1/2}$  & 400 & 1000 & 400 & AMSB \\
$m_{3/2}$ & $10^3$ & $10^3$ & $10^3$ & $5\times 10^4$ \\
$\tan\beta$  & 10 & 10 & 55 & 10 \\
\hline
$m_{\tz_1}$ & 162.9 & 436.3 & 163.8 & 142.1 \\ 
\hline
$\Omega^{std}_{\tz_1} h^2$ & 1.9 & 49.6 & 0.019 & 0.0016 \\
$\sigma^{SI}(\tz_1 p)$ pb & $8.1\times 10^{-10}$ & $1.1\times 10^{-10}$ & 
$2.1\times 10^{-8}$ & $4.3\times 10^{-9}$\\
\hline
\end{tabular}
\caption{Masses and parameters in~GeV units for several benchmark points
computed with \Isajet\,7.81 using $A_0=0$ and $m_t=173.3$ GeV.
}
\label{tab:bm}
\end{table}

In order to keep our results as general as possible, we will not assume particular
PQ parameters, but instead we scan over the following parameter values:
\bea
10^9\ {\rm GeV} < & \fa & < 10^{16}\ {\rm GeV},\label{eq:params1}\\
500\ {\rm GeV}  < & m_{\ta} & < 10^4\ {\rm GeV},\\
10^3\ {\rm GeV} < & m_s & < 10^{5}\ {\rm GeV},\\
0.1 < & \theta_s & < 10,\\
10^5\ {\rm GeV} < & T_R & <10^{12}\ {\rm GeV}\label{eq:params5} .
\eea
Since we will be mostly concerned with the neutralino relic abundance, we leave
the axion mis-alignment angle $\theta_i$ undetermined for now.
\subsection{Benchmark BM1}
\label{ssec:bm1}

\begin{figure}[t]
\begin{center}
\includegraphics[width=12cm]{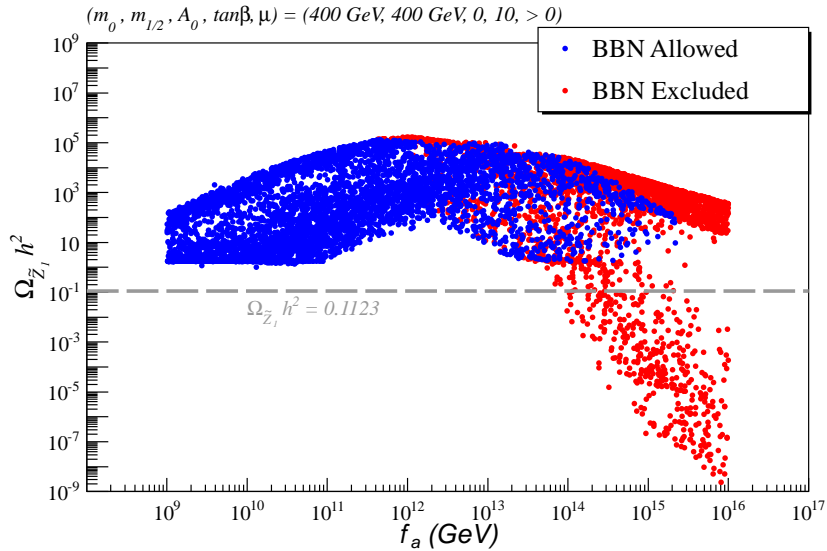}
\caption{Calculated neutralino relic abundance from
mSUGRA model  BM1 versus $\fa$.
We take $m_{\tG}=1$ TeV.  The spread in dots is due to a scan over
PQ parameters $\fa$, $T_R$, $m_{\ta}$, $m_s$ and $\theta_s$.
}
\label{fig:BM1}
\end{center}
\end{figure}

Our results are shown as the resultant relic density of neutralinos $\Omega_{\tz_1}h^2$
in the PQMSSM, where we plot each model versus $\fa$ in Fig. \ref{fig:BM1}. 
The blue points are labeled as BBN-allowed, while red points violate the BBN
bounds as described in Sec. \ref{ssec:bbn}. 
%
%

From Fig.  \ref{fig:BM1}, we see that at low values of
PQ breaking scale $\fa \sim 10^9-10^{11}$ GeV, the value of $\Omega_{\tz_1}h^2$ is 
always bounded from below by its standard value $\Omega_{\tz_1}^{std}h^2\sim 1.9$.
Those points with $\Omega_{\tz_1}h^2\simeq 1.9$ are typically those for which axinos
and saxions decay before $T_{fr}$, or those for which axino/saxion production is 
suppressed by low $T_R$ so that axinos/saxions decays do not significantly contribute to
$\Omega_{\tz_1}h^2$.

In Fig.  \ref{fig:BM1}, frequently the neutralino abundance is enhanced beyond 1.9, 
making these points even more excluded.
The reason why points only have enhanced relic densities at the lower $\fa$ range 
is because the axino-matter coupling is proportional to $1/\fa$, 
and so thermal axino production is enhanced compared to higher $\fa$ values. 
In addition, $\ta\to \tg g$ decays may be phase space suppressed, so that axino decay takes place
at temperature $T_D^{\ta}<T_{fr}$, thereby augmenting the neutralino abundance.
Saxion decay is never phase space suppressed, since $s\to gg$ is always possible, so at the lower
range of $\fa$, saxion decay typically takes place at $T_D^{s}>T_{fr}$.
For values of $\fa \sim 10^{11}$ GeV and beyond, axinos can no longer decay before neutralino 
freeze-out, and so the neutralino abundance is {\it always} enhanced. 
The value of $\fa$ where the neutralino abundance is always enhanced is somewhat 
an artifact of our scanning range, since if we allow $m_{\ta}>10^4$ GeV, 
axinos could become shorter-lived for a higher value of $\fa$ since 
$\Gamma_{\ta}\sim m_{\ta}^3/\fa^2$.

At even higher values of $\fa \agt 10^{12}$ GeV, axino/saxion thermal production becomes 
increasingly suppressed, while saxion production via CO becomes enhanced: entropy dilution
by saxions starts winning over neutralino production from thermal axino production and decay.
Also, both saxion and axino become even longer-lived, so more points become BBN-disallowed.
Although the entropy injection from $s \to gg$ decays grows with $\fa$,
the BBN-allowed blue points are never pushed below $\Omega_{\tz_1}h^2\sim 1.9$, since
$s\to\tg\tg$ also injects additional neutralinos into the thermal bath.\footnote{
We checked the effect of artificially turning off $s\to\tg\tg$ decays in Fig. \ref{fig:BM1}.
In this case, at high $\fa\agt 10^{14}$ GeV, the enhanced saxion production via COs produces only entropy
dilution of the neutralino abundance, and some BBN-allowed points remain with 
a highly suppressed neutralino abundance at high $\fa$. 
This effect has lead to claims that large $\fa\sim M_{GUT}$ values may be allowed
in SUSY models due to entropy injection by saxions\cite{entropy,kim1991,kmy,cck,dine,thomas,acharya,hasen,kkn,bl}.
By properly including $s\to \tg\tg$ decay and the concommitant neutralino re-annihilation at $T_s$, 
the pure entropy injection effect is counter-balanced in this case, and the highly diluted cases 
become BBN-forbidden.
}

\begin{figure}
\begin{center}
\includegraphics[width=14cm]{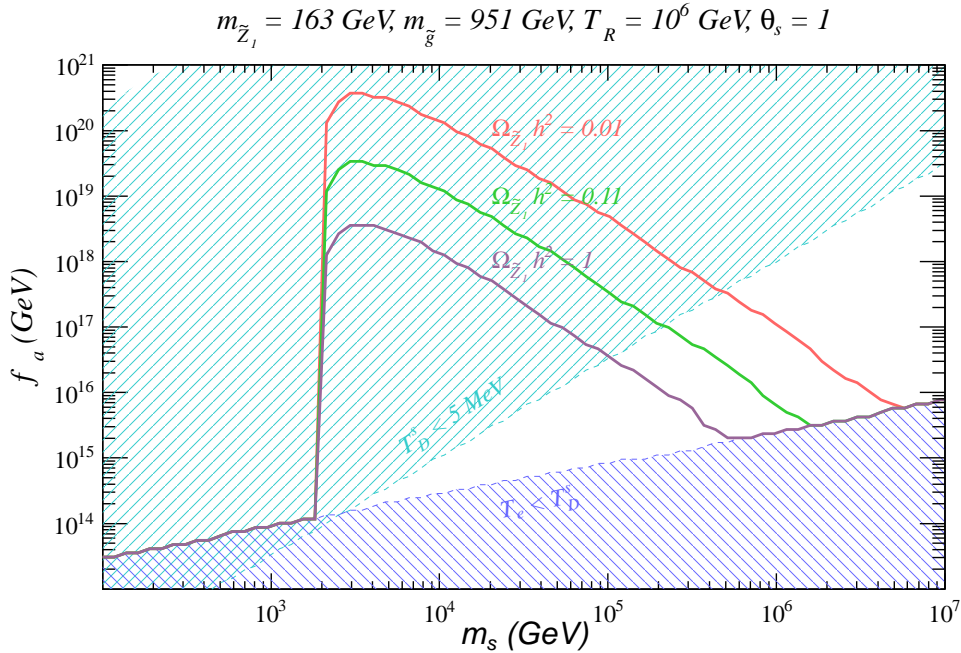}
\caption{Regions in the $m_s\ vs.\ \fa$ plane where $T_D^{s} < 5$~MeV and $T_D^{s} > T_e$. We also show
contours of constant $\Omega_{\tz_1}^{s} h^2$ as estimated using Eq.~\protect \ref{eqap}.}
\label{fig:plane}
\end{center}
\end{figure}

\begin{figure}
\begin{center}
\includegraphics[width=14cm]{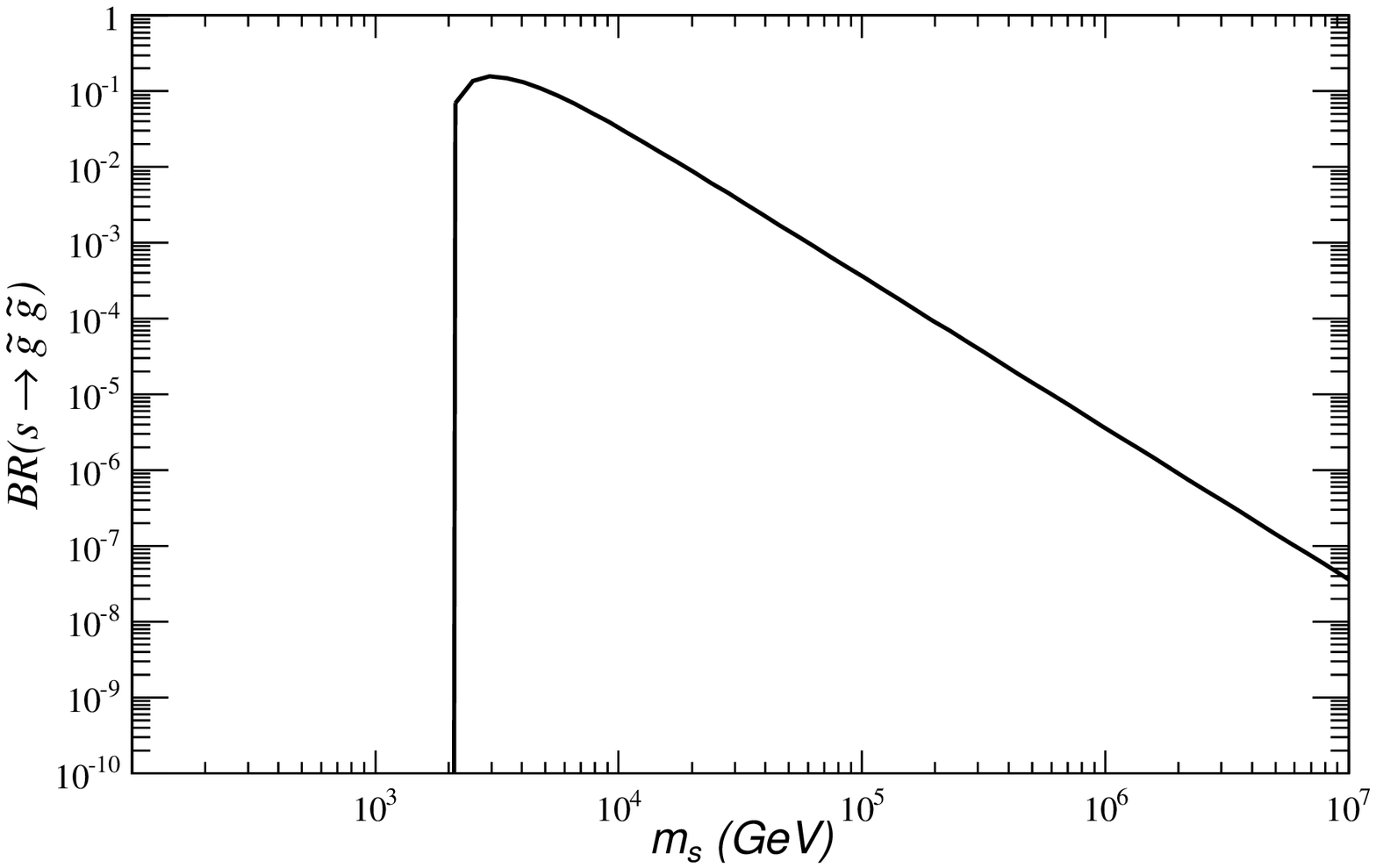}
\caption{Branching ratio of saxion decays into gluino pairs as a function of $m_s$, 
for $m_{\tg} = 951$ GeV.}
\label{fig:BRvsms}
\end{center}
\end{figure}

To understand why the neutralino injection from saxion decays always wins over
the entropy dilution of the neutralino abundance,
we must look at the neutralino Yield from saxion decays.
For simplicity, we will neglect the axino component
as well as neutralino re-annihilation at $T_D^{s}$
and assume that the PQ parameters are chosen so the universe has a saxion-dominated era,
since this is the only scenario with significant entropy injection.
Under these assumptions, the Yield of neutralinos emitted from saxion decays is simply
given by\cite{bl}:
\be
Y_{\tz_1} =  \frac{1}{r} Y_{s} \times 2 BR(s \to \tg\tg)
\ee
where the factor 2 above takes care of the multiplicity of neutralinos from each saxion
decay and $r$ is the entropy injection factor, which can be approximated by
\be
r = \frac{T_e}{T_D^{s}} \; ,
\ee
where $T_D^{s}$ is the saxion decay temperature and $T_e = 4 m_s Y_s/3$ (see Refs.~\cite{bl, ay}). Therefore:
\be
Y_{\tz_1} =  \frac{3}{2} \frac{T_D^{s}}{m_s} BR(s \to \tg\tg) \To \ \ \ 
\Omega_{\tz_1}^{s} h^2 \simeq 4\times 10^8 \mbox{ GeV$^{-1}$ } 
m_{\tz_1} \frac{T_D^{s}}{m_s} BR(s \to \tg\tg) .
\label{eqap}
\ee
The above expression shows that the relic density can be suppressed for large $m_s$, small $T_D^{s}$
and/or small $BR$. However, as seen in Fig. \ref{fig:BM1}, such suppression never seems to drive
$\Omega_{\tz_1} h^2$ below its standard value, except in the BBN excluded region.
To see why this happens, using Eq.~\ref{eqap} we show in Fig. \ref{fig:plane}
contours of $\Omega_{\tz_1}^{s} h^2$ in the $m_s\ vs.\ \fa$ plane for the BM1 point,
with $T_R = 10^6$~GeV and $\theta_s = 1$. We also show the BBN excluded region 
($T_D^{s} < 5$~MeV) and the region with $T_D^{s} > T_e$ ($r < 1$), where there
is no saxion dominated era and Eq.~\ref{eqap} is no longer valid.
As we can see, the allowed region (white) can only satisfy the WMAP constraints at
very large $m_s$ and $\fa$ values. The main reason for the low $\Omega_{\tz_1} h^2$
values obtained in this region is due to the suppression of $BR(s \to \tg\tg)$.
This can be seen in Fig. \ref{fig:BRvsms}, where we show the branching ratio
as a function of $m_s$ for the same benchmark point.
We can see that-- for the region where the $s\to\tg\tg$ decay mode is closed-- the saxion lifetime 
falls into the BBN-forbidden zone. This can also be seen in Fig. \ref{fig:BM1}, where all the
low $\Omega_{\tz_1} h^2$ points at large $\fa$ are BBN excluded.

%

\begin{figure}
\begin{center}
\includegraphics[width=14cm]{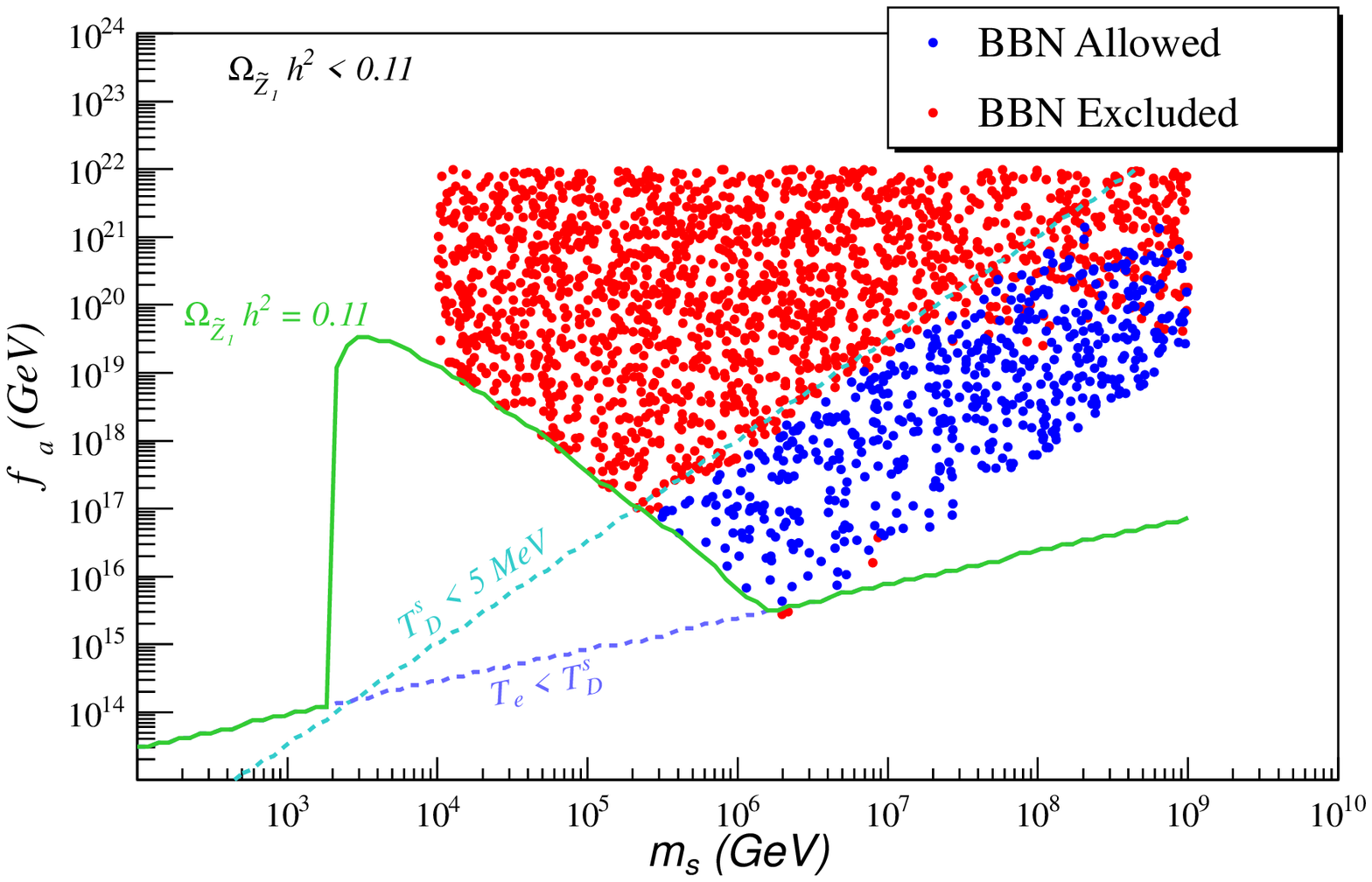}
\caption{Points with $\Omega_{\tz_1} h^2 < 0.11$ obtained through the random
scan described in the text. For comparison, we also show the curves for
$T_D^{s} < 5$~MeV and $T_D^{s} > T_e$ and $\Omega_{\tz_1}^{s} h^2 = 0.11$ 
obtained from Eq.~{\protect \ref{eqap}}.}
\label{fig:scan}
\end{center}
\end{figure}

As seen from the above results, the neutralino relic abundance
can indeed be diluted by including the saxion field, but only at the
expense of going to extremely high $m_s$ and $\fa$ values.
However, since $m_s$ is expected to be of order the {\it soft}
SUSY masses (or $\sim m_{\tG}$ in {\it e.g.} AMSB models), we consider such high values extremely
unnatural. Furthermore, the PQMSSM is most likely not the correct effective
theory at $\fa > 10^{16}-10^{19}$~GeV, where we expect a Grand Unified theory
and/or large supergravity corrections. Nonetheless, to confirm the approximate
results obtained from Eq.~\ref{eqap}, we extend our previous scan over
to 
\be
\fa  \in [10^{15},\;10^{22}]\ {\rm GeV}\;,\ \ \  m_s \in [10^{4},\;10^{9}]\ {\rm GeV}
\ee
and use the full set of Boltzmann equations to compute the neutralino relic abundance.
The results are shown in Fig. \ref{fig:scan}, where we plot in the $m_s\ vs.\ \fa$
plane all solutions satisfying  $\Omega_{\tz_1} h^2 < 0.11$.
As we can see, the numerical results agree very well with the analytical
results in Fig. \ref{fig:plane}. The only discrepancy is in the region near $T_e = T_D^{s}$, 
which does not present viable solutions in the scan. 
This is simply due to the fact that in our estimate of Eq.~\ref{eqap}, 
we neglected the neutralino freeze-out component, 
which becomes dominant when $r \simeq 1$ or $T_e \simeq T_D^{s}$,
increasing the value of $\Omega_{\tz_1} h^2$ in this region.

From the results presented above, we see that {\it for reasonable values
of $\fa$, $m_s$ and $\theta_s$}, the result illustrated for point BM1 in Fig. \ref{fig:BM1} seems to generalize to
all SUSY model points with a standard neutralino overabundance: {\it SUSY models with a standard
overabundance of neutralino dark matter are still at least as excluded when augmented by the PQ mechanism}.

\begin{figure}[t]
\begin{center}
\includegraphics[width=14cm]{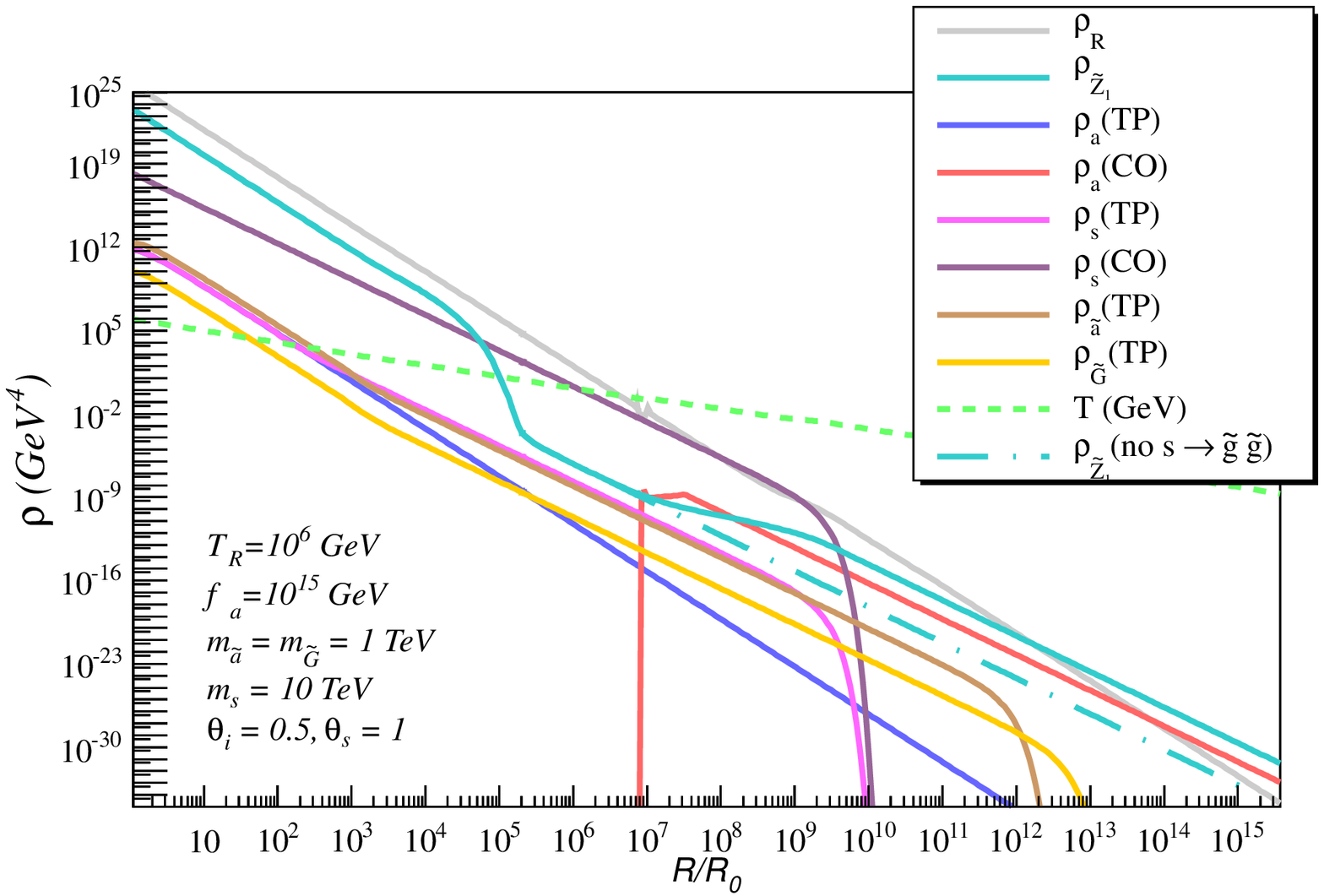}
\caption{Evolution of radiation, neutralino, axion, saxion, axino and gravitino
energy densities versus scale factor $R$ starting at $T=T_{R}$.
We adopt an mSUGRA SUSY model with parameters
$(m_0,m_{1/2},A_0,\tan\beta,sign(\mu ))=(400\ {\rm GeV},400\ {\rm GeV}, 0,10,+)$.
The PQ parameters are listed on the plot.
}
\label{fig:gluinoeffect}
\end{center}
\end{figure}

In Fig. \ref{fig:gluinoeffect}, we show the evolution 
of various energy densities versus the scale factor
for a large $\fa$ value.
In this case, we see that the universe is radiation-dominated out to $R/R_0\sim10^8$, whereupon it
becomes saxion dominated. If only $s\to gg$ is considered, the saxion entropy
injection would cause a large dilution of neutralinos. But by including $s\to\tg\tg$ decays, we
see the neutralino enhancement during $10^7 \lesssim R/R_0 \lesssim 10^9$.
We also show by the dash-dotted line the
expected neutralino energy density by neglecting $s\to \tg\tg$ decays: in this case, the
neutralino abundance is highly suppressed compared to the case where $s\to\tg\tg$ is accounted for.
%

\subsection{Benchmark BM2}
\label{ssec:BM2}

\begin{figure}[t]
\begin{center}
\includegraphics[width=12cm]{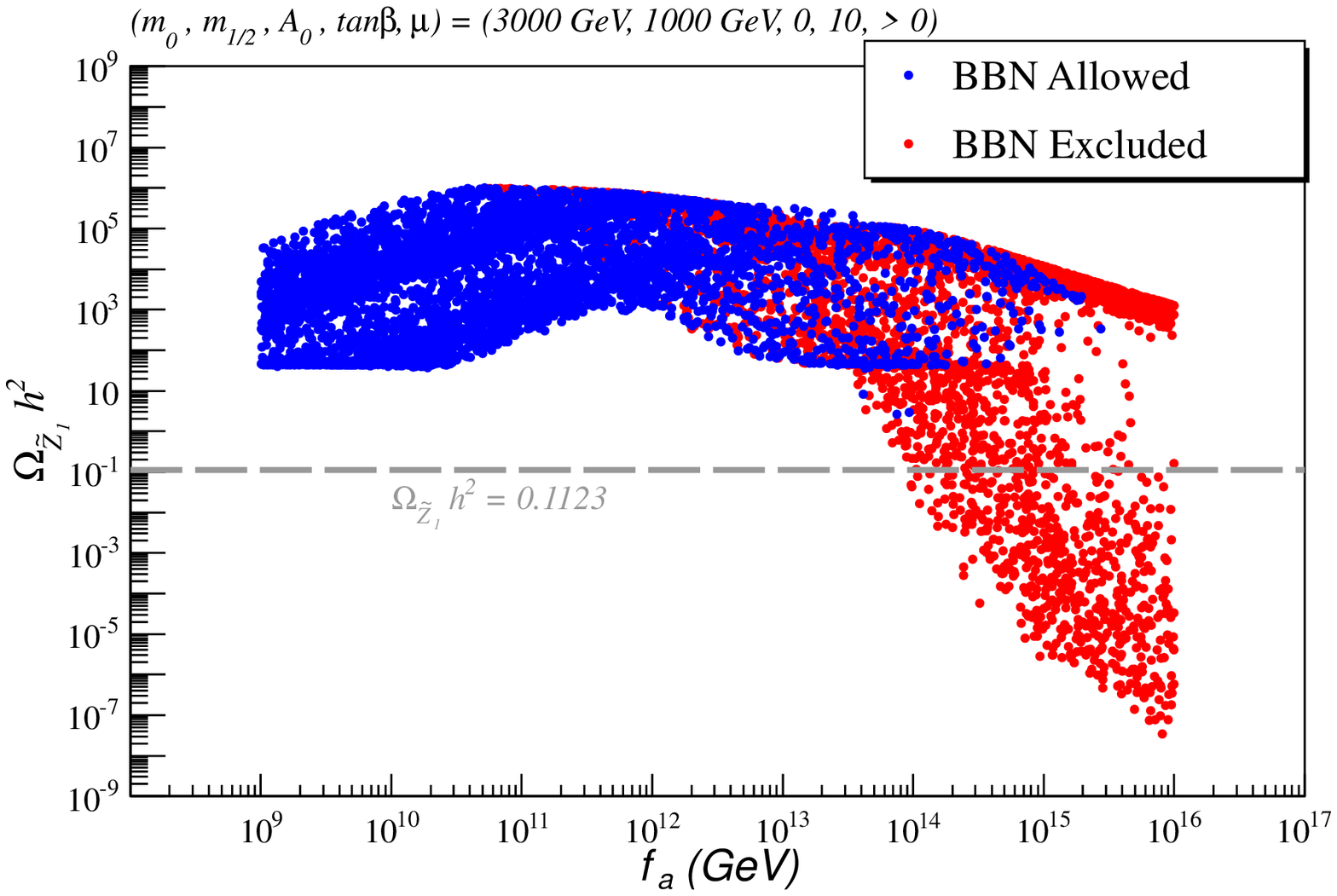}
\caption{Calculated neutralino relic abundance versus $\fa$ from
mSUGRA SUSY model BM2.
The spread in dots is due to a scan over
PQ parameters $\fa$, $T_R$, $m_{\ta}$, $m_s$, $\theta_s$.
}
\label{fig:BM2}
\end{center}
\end{figure}

To emphasize some of the generality of our previous results, we show a further point
with a standard overabundance of neutralinos in Fig.  \ref{fig:BM2}, with 
$(m_0,m_{1/2},A_0,\tan\beta,sign(\mu ))=(3000\ {\rm GeV},1000\ {\rm GeV}, 0,10,+)$,
for which $\Omega_{\tz_1}^{std}h^2\sim 50$. By scanning over PQ parameters, 
again we find that for low $\fa$, $\Omega_{\tz_1}h^2$ either remains at its
standard value (if axinos/saxions decay before freeze-out), or are enhanced (if axinos/saxions
decay after freeze-out). At high $\fa$, entropy dilution from CO-produced saxions
again can suppress $\Omega_{\tz_1}h^2$, but the suppression is counterbalanced
by $s\to\tg\tg$ decays: only BBN-forbidden points where $m_s$ is so light
that $s\to\tg\tg$ is kinematically closed yield points with $\Omega_{\tz_1}h^2<0.11$.
%
%

\subsection{Benchmark BM3: $A$-resonance region}
\label{ssec:BM3}

\begin{figure}[t]
\begin{center}
\includegraphics[width=12cm]{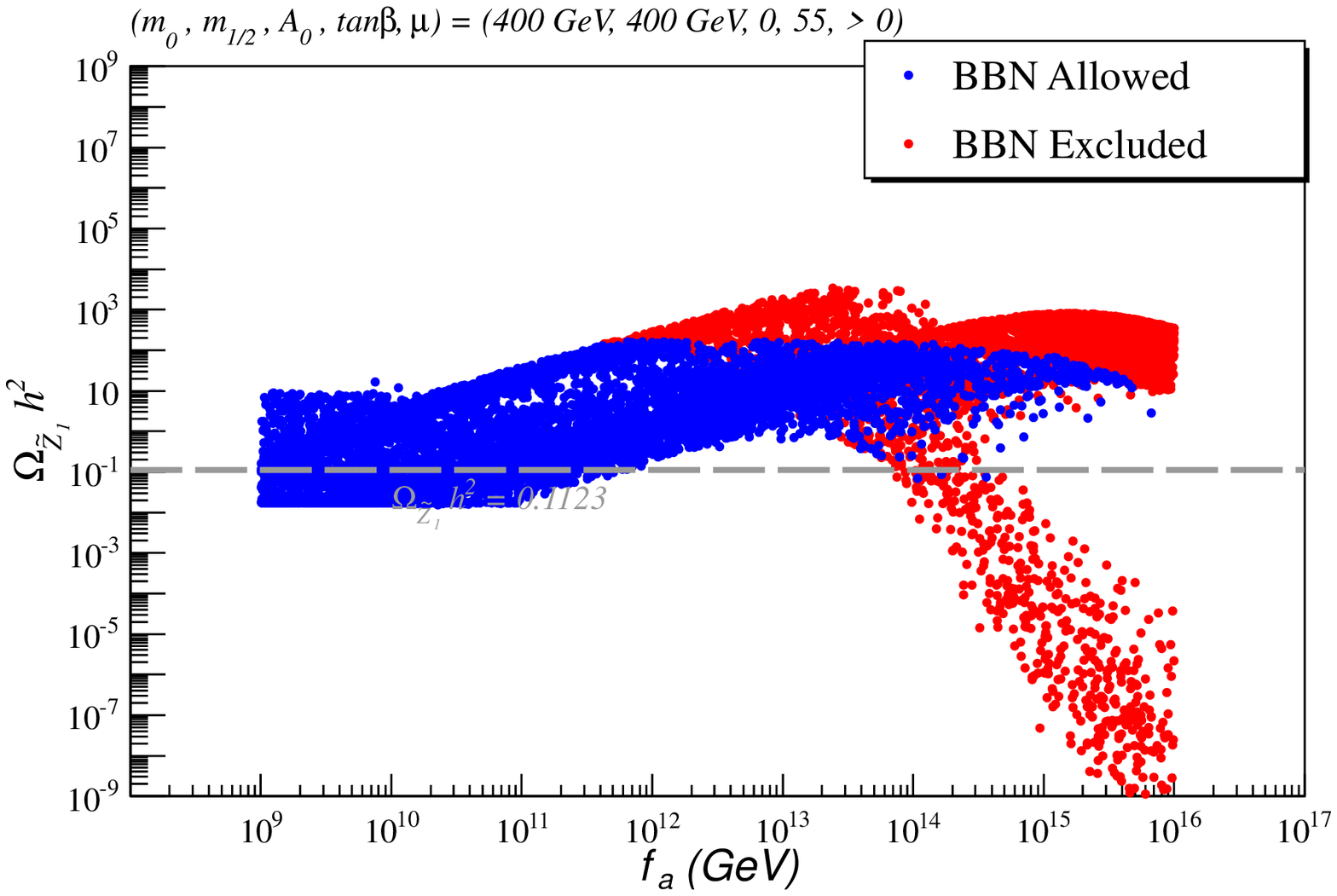}
\caption{Calculated neutralino relic abundance versus $\fa$ from
mSUGRA SUSY model BM3.
The spread in dots is due to a scan over
PQ parameters $\fa$, $T_R$, $m_{\ta}$, $m_s$, $\theta_s$.
}
\label{fig:BM3}
\end{center}
\end{figure}

In Fig.  \ref{fig:BM3}, we show the neutralino abundance in the
case of an mSUGRA point lying in the $A$-resonance annihilation region\cite{Afunnel} 
where $2m_{\tz_1}\sim m_A$. 
We adopt mSUGRA parameters $(m_0,m_{1/2},A_0,\tan\beta,sign(\mu ))=(400\ {\rm GeV},400\ {\rm GeV}, 0,55,+)$,
for which $\Omega_{\tz_1}^{std}h^2\sim 0.02$, {\it i.e.} a standard underabundance.\footnote{
We have also scanned a benchmark point in the hyperbolic branch/focus point region\cite{hb_fp} of mSUGRA,
again with a standard underabundance of neutralino dark matter. The $\Omega_{\tz_1}h^2\ vs.\ \fa$
results look qualitatively much like Fig. \ref{fig:BM3}. 
We do not present these results here in the interests of brevity.
}
In this case, a scan over PQ parameters yields many points at low $\fa$ with 
$\Omega_{\tz_1}h^2\sim 0.02-10$. Thus, the standard neutralino underabundance may be enhanced
up to the WMAP-allowed value, or even beyond. 
As we push to higher $\fa$ values, the axino becomes so long-lived that it only decays after 
neutralino freeze-out, and hence the neutralino abundance is always enhanced. Above $\fa\sim 10^{12}$ GeV, 
the neutralino abundance is enhanced into  the WMAP-forbidden region, with $\Omega_{\tz_1}h^2$
always larger than $0.11$. As we push even higher in $\fa$, then axino production is suppressed, 
but CO-production of saxions becomes large. Entropy dilution turns the range of $\Omega_{\tz_1}h^2$
back down again, and at $\fa\sim 10^{14}$ GeV, some BBN-allowed points again reach 
$\Omega_{\tz_1}h^2\sim 0.11$. In this case, rather large $\fa$ values approaching $M_{GUT}$
are allowed.

%
%

%
\begin{figure}[t]
\begin{center}
\includegraphics[width=12cm]{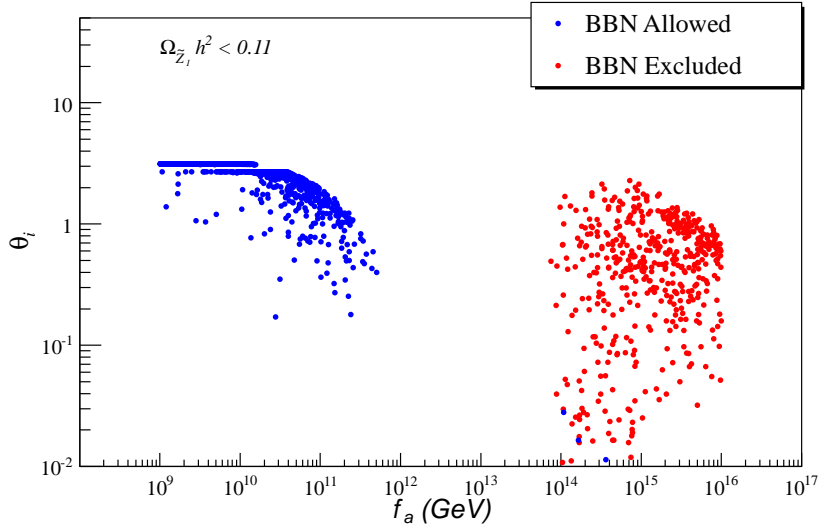}
\caption{Values of the axion mis-alignment angle $\theta_i$
for the points in Fig. {\protect \ref{fig:BM3}} with $\Omega_{\tz_1}h^2<0.11$.
The parameter $\theta_i$ is chosen such as $\Omega_{\tz_1}h^2 + \Omega_{a} h^2 = 0.11$.}
\label{fig:thetaBM3}
\end{center}
\end{figure}
%

For the points with $\Omega_{\tz_1}h^2<0.11$, the 
remaining dark matter abundance can be accommodated by axions via a suitable adjustment of 
the initial axion mis-alignment angle $\theta_i$. In Fig.~\ref{fig:thetaBM3}, we
show the required value of $\theta_i$ needed to enforce $\Omega_{\tz_1} h^2 + \Omega_a h^2 = 0.11$.
At low $\fa$, the points satisfying $\Omega_{\tz_1}h^2<0.11$ have axinos and saxions
decaying before the neutralino freezes out and consequently before axions start to oscillate. 
Hence the axion relic density is not affected by the entropy injection of axinos/saxions and is
 given by the standard expression\cite{vg1}:
\be
\Omega_a h^2\simeq 0.23 \theta_i^2 \left(\frac{\fa}{10^{12}\ {\rm GeV}}\right)^{7/6} \; .
\label{eq:mis-alignment}
\ee
From the above equation, we see that as $\fa$ increases, $\theta_i$ must decrease in order
to maintain $\Omega_{\tz_1} h^2 + \Omega_a h^2 = 0.11$. This behavior is clearly seen
in Fig. \ref{fig:thetaBM3} for $\fa < 10^{12}$~GeV.
Once $\fa$ becomes sufficiently large so axinos and saxions
start to decay after the axion starts to oscillate, the entropy injected from saxions and axinos
considerably dilute the axion relic density, thus allowing for larger $\theta_i$ values.
However, as seen in Fig. \ref{fig:thetaBM3}, this only happens for the BBN-forbidden solutions
at $\fa \gtrsim 10^{14}$~GeV.
The only BBN-allowed points at large $\fa$ with $\Omega_{\tz_1}h^2<0.11$ are the ones
where the saxion production is either suppressed or where it decays before neutralino
freeze-out. In this case there is no significant entropy injection and the axion relic density
is once again given by Eq.~\ref{eq:mis-alignment}. Thus, extremely small values
of $\theta_i$ are required in order to suppress the axion relic density at large $\fa$,
as seen in Fig. \ref{fig:thetaBM3}.
Therefore these points tend to have neutralino domination of the dark matter density, rather than
axion domination. For these points, the large neutralino halo-annihilation rates, enhanced by the
$A$-resonance, may lead to visible production rates of $\gamma$s, $e^+$s and $\bar{p}$s 
in cosmic ray detectors\cite{ofarrill}, while corresponding direct neutralino
detection rates may remain low.

\subsection{Benchmark BM4: AMSB with wino-like neutralino}
\label{ssec:BM4}

\begin{figure}[t]
\begin{center}
\includegraphics[width=12cm]{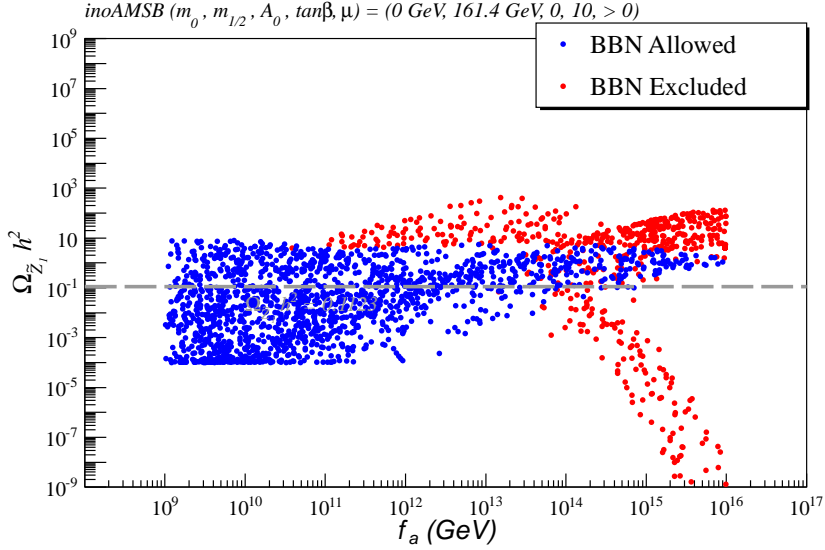}
\caption{Calculated neutralino relic abundance versus $\fa$ from
inoAMSB model BM4.
The spread in dots is due to a scan over
PQ parameters $\fa$, $T_R$, $m_{\ta}$, $m_s$, $\theta_s$.
}
\label{fig:AMSB}
\end{center}
\end{figure}

In Fig.  \ref{fig:AMSB}, we plot $\Omega_{\tz_1}h^2$ for an anomaly-mediated
SUSY breaking model (AMSB) with a wino-like neutralino\cite{amsb}. 
We choose the gaugino-AMSB model with $m_0\sim A_0\sim 0$, 
since this model avoids tachyonic sleptons
without introduction of an additional scalar mass parameter\cite{inoAMSB}.
Model parameters are $(m_{3/2},\ \tan\beta,\ sign(\mu ))=(50\ {\rm TeV}, 10,+)$,
with a standard abundance $\Omega_{\tz_1}^{std}h^2\simeq 0.0016$, far below
the measured value. From the figure, we see that for $\fa\sim 10^{9}-10^{15}$ GeV, 
the neutralino abundance can be enhanced and brought into accord with measured values.
For low $\fa$, axino production and decay augments the abundance, while for
high $\fa$, saxion production and decay both augments and dilutes the abundance.
In this case, as with BM3, the standard underabundance of DM can be augmented and brought into accord
with cosmological measurements. 
Unlike BM3, there exists no intermediate range of $\fa$ which is always excluded
by the production of too much neutralino DM.
The PQ scale can be as large as $\fa\sim 10^{15}$ GeV, in accord with expectations from string theory. 
%
%

Originally, Moroi and Randall had proposed augmenting 
the relic wino abundance from AMSB via moduli production and decay\cite{mr,gg,kane}. 
Here, we see that an alternative mechanism introducing the several PQMSSM fields can also do the job.
Direct and indirect detection rates for wino-like neutralinos have been
presented in Ref. \cite{shibi}.

\section{The case of very large $\theta_s$ and $m_s<2m_{\tg}$}
\label{ssec:largethetas}

From the results presented in Secs.~\ref{ssec:bm1}-\ref{ssec:BM4},
it seems difficult to suppress the neutralino CDM abundance below the standard neutralino abundance.
This conclusion relies on the fact that the (CO) saxion production and decay are correlated through
the value of the PQ scale, since the saxion field strenght ($s(x) = \theta_s\fa$)-- which sets the amplitude of the coherent oscillations--
 is assumed to be of order $f_a$ ($\theta_s \sim \mathcal{O}(0.1-10)$).
Hence large (CO) saxion production only happens at large $f_a$ values and leads to late decaying saxions, usually violating the BBN bounds.
As also discussed above, the BBN bounds can be avoided if saxions have masses in the multi-TeV range,
but then the $s \to \tg \tg$ decay is kinematically allowed and the injection of neutralinos enhances the CDM abundance.
However, if the saxion field strength ($s(x)$) is not set by the PQ breaking scale, but by a much larger
scale, such as the reduced Planck mass (as suggested in some models\cite{kns}), 
it is possible to envision a large production of coherent oscillating saxions even at small $f_a$ values.
In this scenario, small $f_a$ easily satisfies the BBN bounds, allowing for sub-TeV saxion masses, such as $m_s<2m_{\tg}$.
Thus, assuming $s(x) \gg f_a$ ($\theta_s \gg 1$), it is possible to have large saxion production via coherent oscillations, 
small $f_a$ values and small saxion masses without violating the BBN bounds. 
In this case, if $m_s<2m_{\tg}$, saxion decay leads to large entropy production, but does not inject neutralinos.

\begin{figure}[t]
\begin{center}
\includegraphics[width=14cm]{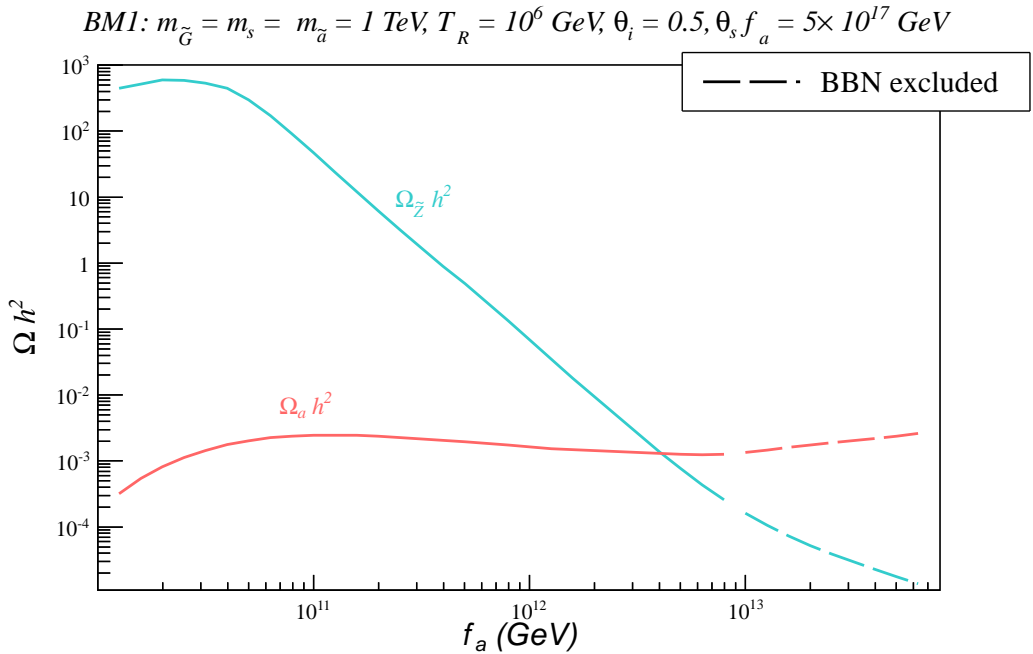}
\caption{Calculated neutralino and axion abundance versus $\fa$ 
for SUSY model BM1 with $\theta_s\fa$ fixed at $5\times 10^{17}$ GeV, and with 
$m_{\tG}=m_s=m_{\ta}=1$ TeV, $T_R=10^6$ GeV and $\theta_i =0.5$.
}
\label{fig:largethetas}
\end{center}
\end{figure}

To illustrate the large $\theta_s$ $(\gg 1)$ scenario, in Fig.~\ref{fig:largethetas} we fix the initial saxion field strength to
$s(x)= \theta_s\fa =5\times 10^{17}$ GeV, but allow $\fa$ to vary and compute the neutralino and axion CDM abundances assuming
$m_{\tG}=m_s=m_{\ta}=1$ TeV, $T_R=10^6$ GeV, $\theta_i =0.5$ and the BM1 benchmark point. In this case, 
$m_s<2m_{\tg}$ so that if saxions can dominate the energy density of the universe, 
they only lead to entropy dilution, and not CDM production. 
From the plot, we see that for low $\fa$ the neutralino abundance is enhanced due
to large thermal production of axinos and their decay to neutralinos.
As $\fa$ increases, thermal production of axinos and saxions becomes 
suppressed, while the saxion decay temperature decreases, leading to increased entropy
dilution of the neutralino abundance. At $\fa\sim 10^{12}$ GeV, $\Omega_{\tz_1}h^2$ drops
below 0.1, and the BM1 point becomes allowed in the PQMSSM. Meanwhile, the axion abundance
will also suffer entropy dilution as $\fa$ increases,
 but this is counterbalanced by an increasing axion field strength, which leads to greater
axion production via COs: the net result is an almost flat value of $\Omega_ah^2$ as
$\fa$ varies. Once $\fa$ increases past $\sim 10^{13}$ GeV, the saxion becomes 
sufficiently long-lived that the model begins to violate BBN bounds. 
While this scenario does provide a strong dilution of dark matter relics, we note here that
the values of $\theta_s$ needed are in the range $\theta_s\sim 10^5-10^6$ so that the
saxion field strength is far beyond the value of $\fa$ and must be given by another physics scale.
%

\section{Conclusions}
\label{sec:conclude}

In this paper we have presented the results of a calculation of mixed
axion/neutralino CDM abundance using a set of eight coupled Boltzmann equations.
The calculation improves upon previous results in several respects: 1. it allows for
non-constant values of $\sigv$, as occurs for bino-like neutralinos, 
where $s$-wave annihilation is suppressed, 2. it allows for interplay between 
neutralino enhancement via axino production and decay, while simultaneously allowing for
neutralino production and dilution via saxion production and decay, 3. it includes
the effect of gravitino production and decay (not a big effect for the parameters 
presented here) and 4. it moves out of the ``sudden decay'' approximation and allows for
continuous axino, saxion and gravitino decay. Our calculation allows for the accurate
estimate of mixed axion/neutralino abundance for general choices of PQMSSM parameters.

In most gravity-mediated SUSY breaking models with gaugino mass unification, 
it is typically the case that the lightest SUSY particle is a bino-like neutralino.
Over most of parameter space of models such as mSUGRA, bino-like neutralinos give rise
to a dark matter abundance far above WMAP limits\cite{bbs2}, 
and hence vast regions of parameter space
are considered as excluded due to overproduction of neutralino dark matter.
In this paper, we have shown that if the MSSM is extended to the PQMSSM-- including
an axion/saxion/axino supermultiplet-- then SUSY models with a standard overabundance of
neutralinos are typically {\it still excluded}, even for very large values of
$\fa\alt 10^{14}-10^{15}$ GeV, 
where it might be expected that a high rate of entropy production from saxion
decay would dilute the DM abundance. Here, we find that $s\to\tg\tg$ compensates against
entropy dilution, and prevents the neutralino abundance from dropping into the measured range, 
unless the saxion decays are in violation of BBN bounds on late-decaying neutral particles.
As noted earlier, our conclusion depends on at least three assumptions. 
First, we implemented the standard thermal axino production rates 
as calculated in the Ref's\cite{ckkr,graf,strumia}. 
These rates should apply in supersymmetric versions of the KSVZ model where 
$PQ$-charged matter multiplets $\hat{\Phi}$ exist at or around the PQ breaking scale $\fa$. 
In a recent publication\cite{cckr}, it has been shown that in the SUSY DFSZ model, 
thermal axino production rates can be enhanced or diminished compared to their KSVZ values
depending on PQMSSM parameters.
Secondly, we assumed that saxion decay is dominated by two-body modes into 
gluon and gluino pairs. In the DFSZ model, decays into Higgs pairs or $aa$ may also
contribute, and even dominate the saxion decay modes.
Thirdly, we have assumed saxion field strength 
$s(x)\equiv \theta_s\fa$ is of order the PQ-breaking scale $\fa$, {\it i.e.} that $\theta_s\sim 1$.
We have also shown in Sec. \ref{ssec:largethetas} that if $\theta_s\gg 1$ and $m_s<2m_{\tg}$, then
CO-produced saxions can dominate the universe and dilute all thermal relics 
while avoiding BBN constraints.

In the case of a standard  underabundance of neutralino CDM, a wide range
of $\fa$ values are permitted, and can augment the neutralino DM into the measured range. 
In cases where the neutralinos still maintain an underabundance, 
the remaining abundance can be accommodated by axions.
In these cases of a standard underabundance of neutralino DM, the PQ scale $\fa$ can be pushed
into the $10^{14}-10^{15}$ GeV range, which is closer to expectations from string theory.
For the case of very high $\fa$, then we typically expect the DM to be neutralino rather than
axion dominated, since the neutralino abundance cannot be suppressed too much without
violating BBN constraints.

\acknowledgments

This research was supported in part by the U.S. Department of Energy,
by the Fulbright Program, CAPES and FAPESP.

\appendix
\section{Boltzmann Equations for the PQMSSM}
\label{sec:boltzeqs}

As discussed in Sec. \ref{sec:calc}, we assume the following set of coupled differential equations:
\bea
\dot{n}_i & = & -  3 H n_i - \Gamma_i m_i \frac{n_i^2}{\rho_i} + [(n^{eq}_{i}(T))^2 - n_{i}^2] \sigv_i + \sum_{j} BR(j,i) \Gamma_j m_j \frac{n_j^2}{\rho_j} \; , \nonumber \\
\dot{S} & = &\frac{R^3}{T} \sum_{i} BR(i,X) \Gamma_i m_i n_{i}  \; ,
\label{diffs}
\eea
with $H$ given by:
\be
H = \frac{1}{R} \frac{d R}{d t} = \sqrt{\frac{\rho_T}{3 M_P^2}} \; , \label{H2}
\ee
where $\rho_T$ is the total energy density.

In order to simplify the above equations we define:
\be
x = \ln(R/R_0),\;\; N_i = \ln(n_i/s_0),\;\; {\rm and}\;\; N_S = \ln(S/S_0)
\ee
so we can write Eq's.~\ref{diffs} as:
\bea
N_S' & = & \frac{1}{HT} \sum_{i} BR(i,X) \Gamma_i m_i \exp[N_i + 3 x - N_S] 
\label{RSeq} \\
N_i' & = & - 3 - \frac{\Gamma_i}{H} \frac{m_i}{\rho_i/n_i} + \sum_{j\neq i} BR(j,i) \frac{\Gamma_j}{H} \frac{m_j}{\rho_j/n_j} \frac{n_j}{n_i} + \frac{\sigv_i}{H} n_i [\left(\frac{n_i^{eq}}{n_i}\right)^2 -1] \label{Nieq}
\eea
where $'=d/dx$ and $n_i$ is given by $n_i = s_0 e^{N_i}$.

The above equation for $N_i$ also applies for coherent oscillating fields, if we define:
\be
N_i = \ln(n_i/s_0),\;\; {\rm and}\;\; n_i \equiv \rho_i/m_i
\ee
so
\be
N_i' = -3 - \frac{\Gamma_i}{H} \label{Nico}
\ee
where we assume that the coherent oscillating component does not couple to any of the other fields.

Since $H$ depends on the energy densities, to solve the above equations 
we must compute $\rho_i$ from $n_i$. However, even for particles following a thermal distribution,
the energy density for each component cannot be directly obtained from $n_i$, 
unless the chemical potential ($\mu_i$) is also given.
Nonetheless, $\mu_i(T)$ is usually small in the relativistic regime, while in the non-relativistic regime
we always have $\rho_i = m_i n_i$. Therefore, {\it assuming that the 
fields follow a thermal distribution}, a good approximation for $\rho_i$ as a function of $n_i$ is given by:
\be
\rho_i = n_i \times \left\{ 
\begin{array}{ll} 
m_i & , \mbox{ if $T_i < m_i/10$} \\
m_i \frac{K_1(m_i/T_i)}{K_2(m_i/T_i)} + 3 T_i & , \mbox{ if $m_i/10 < T_i < 3 m_i/2$} \\
N_F \frac{\pi^4}{\xi(3)} \frac{T_i}{30} & , \mbox{ if $3 m_i/2 < T_i$}
\end{array} \right. \label{rhoi}
\ee
where the modified Bessel functions, $K_1$ and $K_2$, are necessary to describe a smooth relativistic/non-relativistic transition and $N_F = 1 (7/6)$ for bosons (fermions).

The only remaining piece of information necessary for computing $\rho_i$ and $H$ and solving the Boltzmann
equations is the definition of temperature for each component. The radiation temperature can be directly obtained from $N_S$ and $x$:
\be
T = \left(\frac{g_*(T_R)}{g_*(T)}\right)^{1/3} T_R \exp[N_S/3 -x] .
\ee
For thermal fluids in equilibrium we always have $T_i =T$, but
once they decouple, this is no longer true. However, the
temperature of relativistic fluids scales as $T \propto R^{-1}$, 
while non-relativistic fluids have $T\propto R^{-2}$.
Thus, we approximate $T_i$ by
\be
T_i = \times \left\{ 
\begin{array}{ll} 
T &  , \mbox{ if coupled} \\
T_i^{dec} \frac{R_i^{dec}}{R} & , \mbox{ if $T_i > 3m_i/2$ and decoupled} \\
\frac{3}{2} m_i \left(\frac{R_i^{NR}}{R}\right)^2 & , \mbox{ if $T_i < 3m_i/2$ and decoupled} 
\end{array} \right. \label{Ti}
\ee
where $T_i^{dec}$, $R_i^{dec}$ and $R_i^{NR}$ are the decoupling (freeze-out) temperature, the scale factor at freeze-out
and the scale factor at the non-relativistic transition ($T_i = 3 m_i/2$), respectively.
If the fluid was never in thermal equilibrium, we take $T_i^{dec}= T_R$.
For coherent oscillating fluids we always have $T_i = 0$\footnote{In principle, the approximations
in Eq's.~\ref{rhoi} and \ref{Ti} can be avoided if we include equations for the chemical potentials $\mu_i(T)$. 
However, for simplicity, we use Eq's.~\ref{rhoi} and \ref{Ti} instead.}.

Eq's.~\ref{RSeq} and \ref{Nieq}, with the auxiliary equations for $H$ (Eq.~\ref{H2}), $\rho_i$ 
(Eq.~\ref{rhoi}) and
$T_i$ (Eq.~\ref{Ti}) form a set of closed equations, which can be solved once the initial conditions for the number
densities ($n_i$) and entropy ($S$) are given. 
The initial entropy $S_0$ is trivially obtained, once we assume a radiation dominated universe
at $T=T_R$:
\be
S(T_R) = \frac{2 \pi^2}{45} g_*(T_R) T_R^3 R_0^3 .
\ee
For thermal fluids we take the initial number density as
\be
n_i(T_R) = \left\{ 
\begin{array}{ll} 
0 & , \mbox{ if $\sigv_i n^{eq}_i/H|_{T=T_R} < 10$} \\
n_i^{eq}(T_R) & , \mbox{ if $\sigv_i n^{eq}_i/H|_{T=T_R} > 10$} 
\end{array} \right. ,
\label{ni0TP}
\ee
while for coherent oscillating fluids the initial condition is set at the beginning of oscillations:
\be
n_i(T^{osc}_i)=\frac{\rho_i^{0}}{m_i(T^{osc}_i)}
\ee
where $T^{osc}_i$ is the oscillation temperature, given by $3H(T^{osc}_i) = m_i(T^{osc}_i)$
and $\rho_i^{0}$ the initial energy density for oscillations.
For the oscillating saxion and axion\cite{vg1} fields the initial energy densities are given by:
\begin{eqnarray*}
\rho^0_a & = & 1.44  \frac{m_a(T)^2 (\fa)^2 \theta_i^2 }{2} f(\theta_i)^{7/6}\\
\rho^0_s & = & \min\left[2.1 \times 10^{-9}\left(\frac{2\pi^2 g_*(T_R) T_R^3}{45}\right)\left(\frac{T_R}{10^5}\right)\left(\frac{\theta_s (\fa)}{10^{12}}\right)^2,\frac{m_s^2 \theta_s^2 (\fa)^2}{2}\right]	
\end{eqnarray*}
where $f(\theta_i) = \ln[e/(1-\theta_i^2/\pi^2)]$ and $\theta_i \fa$ and $\theta_s \fa$ are the initial axion and saxion field amplitudes. 
The definition of $\rho^0_s$ accounts for the possibility of saxion oscillations beginning 
during inflation (if $T_R < T_{osc}$).

In order to compute the source term in Eq.~\ref{diffs}, we must specify the annihilation
cross-sections $\sigv_i$, the branching ratios $BR(i,j)$ and $BR(i,X)$ and the the decay 
widths $\Gamma_i$. 
The annihilation cross-sections for axions, saxions, axinos and gravitinos are given by the 
expressions \cite{graf,strumia,pradler}
\begin{eqnarray*}
\sigv_{a} & = & 10^{-4} \frac{g_s^6}{(\fa)^2} \left[ 4.19 \ln(1.5/g_s^2) + 1.68\times \theta(T - m_a(T))  \right]\\
\sigv_{\ta} & = & 10^{-5} \frac{g_s^6}{(\fa)^2} \left[ 3 + 3.87 \times \theta(T - m_{\ta}) \left\{ \begin{array}{ll} 
23.863 g_s^{-0.7} - 0.784 &, \mbox{if $g_s > 0.35$} \\ 
4.47+ 31 \ln(1.4/g_s) - 0.784 &, \mbox{if $g_s < 0.35$} \end{array} \right. \right]\\
\sigv_{\tG} & = & \frac{1.37}{M_P^2}\times \left[ 72 g_s^2 \ln(1.271/g_s)(1+\frac{M_3^2}{3 m_{\tG}^2})\right. \\
&+& \left. 27 g^2 \ln(1.312/g)(1+\frac{M_2^2}{3 m_{\tG}^2}) + 11 g'^2 \ln(1.266/g')(1+\frac{M_1^2}{3 m_{\tG}^2}) \right] ,
\end{eqnarray*}
while $\langle\sigma v\rangle_{\tz_1}(T)$ is extracted from IsaReD\cite{isared}.
The second term in the expressions for $\sigv_a$ and $\sigv_{\ta}$ represent contributions from $1\to 2$ decays
of particles with thermal masses. Therefore, these terms should not be included unless $T > m_{a,\ta}$, as indicated
by the $\theta$ functions above. The expression for
the axino effective cross-section is set to reproduce the numerical results in \cite{strumia}. 
Since the saxion thermal production has not been computed, we approximate it by the axion expression:
\be
\sigv_{s} = \sigv_{a}
\ee
with $m_a \to m_s$.

For obtaining the various unstable particle widths,
we calculate $\Gamma_{\ta}$ from the $\ta\to\tg g$, $\tz_i\gamma$ and
$\tz_i Z$ partial widths as presented in Ref. \cite{blrs}. For gravitino decays, 
we adopt the gravitino widths as presented in Ref. \cite{moroi_grav}. For the saxion width, 
we include $\Gamma_s$ from the $s\to gg$ and $s\to \tg\tg$ decays as presented in Ref. \cite{ay}.
We note here that in the DFSZ model, it is also possible to have $s\to hh$ decays and
possibly $s\to aa$ decays. We neglect these latter two cases, so that our results apply to
the supersymmetrized KSVZ model, where the $gg$ and $\tg\tg$ final states should dominate.

Once the total and partial widths are known, we can easily compute the required branching ratios:
\be
BR(\ta,\tz_1) = 1,\; BR(s,\tz_1) = 2 \times \frac{\Gamma(s \to \tg \tg)}{\Gamma_s}, \; BR(\tG,\tz_1) = 1
\ee
The factor $2$ in $BR(s,\tz_1)$ takes care of the multiplicity of neutralinos for each saxion
cascade decay. While the $s\to gg$ decay width is always dominant, we showed in
Sec. \ref{sec:numerics} that $s\to\tg\tg$ plays a crucial role in the PQMSSM dark matter cosmology.

Finally, we assume that the branching ratios for computing the energy injection into the
thermal bath from unstable particle decays are given by:
\be
BR(\ta,X) = BR(s,X) = BR(\tG,X) = 1 .
\label{brx}
\ee
Although some of the decay energy is lost into neutralinos (except for $s \to gg$ decays),
 we assume that in the final product of the cascade decay of axinos, saxions and gravitinos 
most of the initial energy has been converted into radiation, so Eq.~\ref{brx} consists in a
good approximation.

%

%
\end{document}